\documentstyle[12pt,fleqn]{article}
\begin{document}

\author{\bf{Shahin S. Agaev}\\ \it High Energy Physics Lab., Baku State
University,\\ \it Z.Khalilov st.23, 370148 Baku, Azerbaijan\thanks{E-mail: 
azhep@lan.ab.az}}
\title{\bf{RESUMMATION OF INFRARED RENORMALONS IN THE PION ELECTROMAGNETIC FORM
FACTOR CALCULATIONS}}
\date{}
\maketitle
\begin{abstract}
The pion electromagnetic form factor $F_{\pi }(Q^{2})$ is calculated at the
leading order of pQCD using the running coupling constant $(\sim \alpha
_{S}(Q^{2}(1-x)(1-y))$ approach. The resummed expression for $F_{\pi
}(Q^{2}) $ is found. It is shown that the effect of the infrared renormalons
can be taken into account by scale-setting procedure $\alpha
_{S}(Q^{2})\rightarrow \alpha _{S}(e^{f(Q^{2})}Q^{2})$ in the leading order
expression.
\end{abstract}
\newpage

{\bf 1.}   It is known that one of the important problems in the
perturbative QCD (pQCD) calculations is a choice of the renormalization
scheme and scale in a truncated perturbative series [1,2]. Of course, when
working to all orders in $\alpha _{S}(Q^{2})$ the choice of scale and scheme
is irrelevant, because a measurable physical quantity {\bf f}is the
same for all choices. But this freedom can be a source of difficulties, when
one truncates a perturbative series for {\bf f}. Namely, a bad choice of
the coupling constant scale may result in large higher-order corrections in
such truncated series. The method proposed by Brodsky, Lepage and Mackenzie
(BLM) in Refs.[1,2] for determination of the coupling constant scale seems
to be a good tool for the handling of these problems. The BLM scale-setting
procedure amounts to absorbing certain vacuum polarization corrections
appearing at higher-order into the one-loop QCD running coupling constant.
In other words, the essence of this method is to find an average virtuality
of a gluon in a loop diagram and to use it as the scale in the running
coupling constant. As was emphasized in [1,2], this method may be applied to
both inclusive and exclusive processes.

Recently, the BLM method has been generalized by performing the calculations
with the running coupling constant $\alpha _{S}(-k^{2})$ at the vertices of
Feynman diagrams, where k is the momentum flowing through the virtual gluon
line [3,4]. This approach is equivalent to a resummation of an infinite
number of terms in the perturbative series for the quantity {\bf f}.

In our previous works [5,6] we have used the running coupling constant
approach for calculation of the pion and kaon electromagnetic form factors
at the leading order of pQCD. As a result we have found the perturbative
series in $\alpha _{S}(Q^{2})$ for $Q^{2}F_{M}(Q^{2})$. In this letter we
shall resum the series obtained in Ref.[5] for the pion form factor using
the technique of the Borel transform and show that an effect of such
resummation can be taken into account by redefining the scale of $\alpha
_{S}(Q^{2})$ in the leading order expression.

\bigskip

{\bf 2.}   Let us start from the factorized QCD expression for the
pion electromagnetic form factor 
\begin{equation}
F_{\pi }(Q^{2})=\int_{0}^{1}\int_{0}^{1}dxdy\phi _{\pi }(y,\widehat{Q}%
^{2})T_{H}(x,y;Q^{2},\alpha _{S}(\widehat{Q}^{2}))\phi _{\pi }(x,\widehat{Q}%
^{2}),  \label{1}
\end{equation}
where $Q^{2}=-q^{2}$ is the four momentum square of the virtual photon $%
\gamma ^{*}$, $\widehat{Q}^{2}$ is the renormalization and factorization
scale. In (1) terms of order $m_{\pi }^{2}/Q^{2}$, as well as higher Fock
state components of the pion wave function have been neglected. Here the
pion wave function $\phi _{\pi }(x,\widehat{Q}^{2})$ is non-perturbative
quantity and gives the amplitude for finding the quark (antiquark) carrying
the longitudinal fractional momentum $x(1-x)$ within the pion. The function $%
T_{H}(x,y;Q^{2},\alpha _{S}(\widehat{Q}^{2}))$ is the hard scattering
amplitude for the subprocess $q\overline{q}^{\prime }+\gamma ^{*}\rightarrow 
$ $q\overline{q}^{\prime }$ and can be found by using pQCD, 
\begin{equation}
T_{H}(x,y;Q^{2},\alpha _{S}(\widehat{Q}^{2}))=\alpha _{S}(\widehat{Q}%
^{2})T_{H}^{1}(x,y,Q^{2})\left[ 1+\alpha _{S}(\widehat{Q}%
^{2})T_{H}^{2}(x,y,Q^{2})+...\right] .  \label{2}
\end{equation}
The leading order term $\alpha _{S}(\widehat{Q}^{2})T_{H}^{1}(x,y,Q^{2})$
here is given by [7] 
\begin{equation}
\alpha _{S}(\widehat{Q}^{2})T_{H}^{1}(x,y,Q^{2})=\frac{16\pi C_{F}\alpha
_{S}(\widehat{Q}^{2})}{Q^{2}(1-x)(1-y)},\qquad C_{F}=4/3.
\end{equation}
The next-to-leading order correction $T_{H}^{2}(x,y,Q^{2})$ was found in
[8]. The one-loop QCD running coupling constant $\alpha _{S}(\widehat{Q}%
^{2}) $ in (2), (3) is defined as 
\begin{equation}
\alpha _{S}(\widehat{Q}^{2})=\frac{4\pi }{\beta _{0}\ln (\widehat{Q}%
^{2}/\Lambda ^{2})},  \label{4}
\end{equation}
where 
\[
\beta _{0}=11-\frac{2}{3}n_{f}, 
\]
and $n_{f}$ being the number of quark flavors.

In Ref.[1] the authors argued that in a meson form factor calculations the
scale in (3) has to be chosen equal to the square of a momentum transfer of
the exchanged gluon $\widehat{Q}^{2}=Q^{2}(1-x)(1-y)$. But it is evident
that $\alpha _{S}(Q^{2}(1-x)(1-y))$ suffers from singularities associated
with the behavior of $\alpha _{S}$ in the soft region $x\rightarrow
1,y\rightarrow 1$. In other words, integrations over $x,y$ in (1) extended
to $x\rightarrow 1,y\rightarrow 1$ indicate the appearance of
non-perturbative effects. Therefore, until now in all pQCD calculations the
scale $\widehat{Q}^{2}$ was taken equal to some fixed quantity (as a rule, $%
\widehat{Q}^{2}\equiv Q^{2}/4$) and results for $F_{\pi }(Q^{2})$ were
obtained in the so-called ''frozen coupling constant'' approximation. In our
previous paper [5] for calculation of $F_{\pi }(Q^{2})$ in the running
coupling constant approach we have expressed $\alpha _{S}(Q^{2}(1-x)(1-y))$
by means of the renormalization group equation in terms of $\alpha
_{S}(Q^{2})$ and calculated obtained integrals over $x,y$ as principal value
ones (see, [9,10]). For $Q^{2}F_{\pi }(Q^{2})$ we have found the following
perturbative series in $\alpha _{S}(Q^{2})$, 
\begin{equation}
Q^{2}F_{\pi }(Q^{2})=\left( 16\pi rf_{\pi }\right) ^{2}\sum_{n=1}^{\infty }
\left( \frac{\alpha _{S}(Q^{2})}{4\pi }\right)
^{n}S_{n},  \label{5}
\end{equation}
where $r$ and $S_{n}$ are the model dependent coefficients. In calculations
we have used the asymptotic $\phi _{asy}(x)$ and the two-humped
Chernyak-Zhitnitsky $\phi _{CZ}(x,\widehat{Q}^{2})$ wave functions [11], 
\begin{eqnarray}
\phi _{asy}(x) &=&\sqrt{3}f_{\pi }x(1-x),  \nonumber \\
\phi _{CZ}(x,\widehat{Q}^{2}) &=&\phi _{asy}(x)\left\{ 1+\frac{2}{3}%
C_{2}^{3/2}\left( 2x-1\right) \left[ \frac{\alpha _{S}(\widehat{Q}^{2})}{%
\alpha _{S}(\mu _{0}^{2})}\right] ^{\gamma _{2}/\beta _{0}}\right\} .
\label{6}
\end{eqnarray}
Here $f_{\pi }=0.093~GeV$ is the pion decay constant, $C_{2}^{3/2}(2x-1)$ is
the Gegenbauer polynomial, $\gamma _{2}=50/9$ is the anomalous dimension, $%
\mu _{0}=0.5~GeV$ is the normalization point. The phenomenological pion wave
function $\phi _{CZ}(x,\widehat{Q}^{2})$ depends on the scale $\widehat{Q}%
^{2}$. But in this article we ignore this dependence and take it equal to $%
\phi _{CZ}(x,\mu _{0}^{2})$. The coefficients $S_{n}$ of the perturbative
series (5) and $r$ for $\phi _{asy}(x)$ and $\phi _{CZ}(x,\mu _{0}^{2})$ are
given by the following expressions, respectively 
\begin{equation}
S_{n}=c_{n}\beta _{0}^{n-1}=(n-1)!\left( n-2+\frac{n+4}{2^{n+1}}\right)
\beta _{0}^{n-1},\qquad r=1,  \label{7}
\end{equation}
and 
\begin{eqnarray}
S_{n}=c_{n}\beta _{0}^{n-1}&=&(n-1)!\left[ -\frac{14}{3}+n\left( 1+\frac{25%
}{2^{n+1}}+\frac{64}{3^{n+1}}+\frac{1}{4^{n-1}}\right) -\right.  \nonumber \\
&&\left. \frac{25}{2^{n-1}}+\frac{8}{3^{n}}+\frac{35}{3\cdot 4^{n-1}}\right]
\beta _{0}^{n-1},\qquad r=5\quad 
\end{eqnarray}

The $S_{n}$ in both cases are proportional to $(n-1)!$, as a result the
series (5), (7), (8) are ill-defined and must be resummed using some
procedures for removing divergences. The procedure for such resummation is
well known; perform the Borel transform in (5) with respect to the inverse
coupling constant (see, for example, [3,12]) and after that calculate the
inverted integral applying some regularization methods.

\bigskip

{\bf3.}   The Borel transform of the series (5) is 
\begin{equation}
B[Q^{2}F](u)=\sum_{n=1}^{\infty }\frac{u^{n-1}}{(n-1)!%
}c_{n}.  \label{9}
\end{equation}
Eq.(9) can be formally inverted to give 
\begin{equation}
\left[ Q^{2}F_{\pi }(Q^{2})\right] ^{res}=\frac{(16\pi rf_{\pi })^{2}}{\beta
_{0}}\int_{0}^{\infty }du\exp \left[ -4\pi u/\beta _{0}\alpha
_{S}(Q^{2})\right] B[Q^{2}F](u).  \label{10}
\end{equation}
In these cases, when one finds integral (10), it defines the Borel sum 
$[Q^{2}F_{\pi}(Q^{2})] ^{res}$ of the series (5). For the
solution of our problem we may follow this general recipe to find $\left[
Q^{2}F_{\pi }(Q^{2})\right] ^{res}.$ But by means of the identity 
\begin{equation}
\int_{0}^{\infty }\exp \left[ -u(t+w+z)\right] =\frac{1}{t+w+z},  \label{11}
\end{equation}
Eq.(1) can be written in a more convenient way [5,6]. Indeed, after
substituting the leading order term in the renormalization group equation's
solution [10] 
\begin{equation}
\alpha _{S}(Q^{2}(1-x)(1-y))\simeq \frac{\alpha _{S}(Q^{2})}{1+\frac{1}{t}%
\ln [(1-x)(1-y)]},\qquad \frac{1}{t}=\frac{\alpha _{S}(Q^{2})\beta _{0}}{%
4\pi },  \label{12}
\end{equation}
into (1), and changing of the variables $x,~y$ to $z=\ln (1-x),~w=\ln (1-y)$,
introducing the formula (11) and integrating over $z,w$ for the asymptotic
wave function we find 
\begin{eqnarray}
Q^{2}F_{\pi }(Q^{2}) &=&\frac{(16\pi f_{\pi })^{2}}{\beta _{0}}%
\int_{0}^{\infty }du\exp \left[ -4\pi u/\beta _{0}\alpha _{S}(Q^{2})\right]
\left[ \frac{1}{(1-u)^{2}}+\right.  \nonumber \\
&&\left. \frac{1}{(2-u)^{2}}-\frac{2}{1-u}+\frac{2}{2-u}\right] .  \label{13}
\end{eqnarray}
It is evident that 
\begin{equation}
B[Q^{2}F]^{asy}(u)=\frac{1}{(1-u)^{2}}+\frac{1}{(2-u)^{2}}-\frac{2}{1-u}+%
\frac{2}{2-u}.  \label{14}
\end{equation}
The Borel transform of the series (5), (8) is 
\begin{eqnarray}
B[Q^{2}F]^{CZ}(u) &=&\frac{1}{(1-u)^{2}}+\frac{25}{(2-u)^{2}}+\frac{64}{%
(3-u)^{2}}+\frac{16}{(4-u)^{2}}-  \nonumber \\
&&\frac{14}{3(1-u)}-\frac{50}{2-u}+\frac{8}{3-u}+\frac{140}{3(4-u)}.
\label{15}
\end{eqnarray}
It is easy by using 
\[
c_{n}=\left( \frac{d}{du}\right) ^{n-1}\left. B[Q^{2}F](u)\right| _{u=0}, 
\]
to recover the series (5), (7), (8).

The Borel transforms (14), (15) have singularities on the real $u$ axis $($%
at $u=1,2,3,4)$, which are known as infrared renormalons. The infrared
renormalons are responsible for a factorial growth of the expansion
coefficients in Eqs.(7),(8). This fact confirms our conclusion in [5]
concerning an infrared renormalon nature of divergences in the perturbative
series (5), (7), (8).

The difficulties with the double and single poles at $u=1,2$ in (13) can be
avoided by taking the integral in its principal value. Then we obtain 
\begin{equation}
\left[ Q^{2}F_{\pi }(Q^{2})\right] _{asy}^{res}=\frac{(16\pi f_{\pi })^{2}}{%
\beta _{0}}\left[ -\frac{3}{2}+(\ln \lambda -2)\frac{Li(\lambda )}{\lambda }%
+(\ln \lambda +2)\frac{Li(\lambda ^{2})}{\lambda ^{2}}\right] .  \label{16}
\end{equation}
The similar calculations lead to 
\begin{eqnarray}
\left[ Q^{2}F_{\pi }(Q^{2})\right] _{CZ}^{res} &=&\frac{(80\pi f_{\pi })^{2}%
}{\beta _{0}}\left[ -\frac{233}{6}+(\ln \lambda -\frac{14}{3})\frac{%
Li(\lambda )}{\lambda }+\right.  \nonumber \\
&&25(\ln \lambda -2)\frac{Li(\lambda ^{2})}{\lambda ^{2}}+8(8\ln \lambda +1)%
\frac{Li(\lambda ^{3})}{\lambda ^{3}}+  \nonumber \\
&&\left. 4(4\ln \lambda +\frac{35}{3})\frac{Li(\lambda ^{4})}{\lambda ^{4}}%
\right] .  \label{17}
\end{eqnarray}
In (16),(17) $Li(\lambda )$ is the logarithmic integral [13] 
\begin{equation}
Li(\lambda )=\int_{0}^{\lambda }\frac{dx}{\ln x},\qquad \lambda
=Q^{2}/\Lambda ^{2}.  \label{18}
\end{equation}

The principal value prescription used above to regulate the infrared
renormalon singularities at $u=1,2,3,4$ in Eq.(10) (or at $x=1$ in (18 ))
produces an ambiguity in the perturbative series for the pion form factor $%
F_{\pi }(Q^{2})$. Any other approach for defining the integral (10) would
differ in the treatment of these singularities. The ambiguity introduced by
our treatment, as has been shown in [5], is a higher twist and for $%
Q^{2}F_{\pi }(Q^{2})$ behaves as $\Lambda ^{2}/Q^{2}$. It is well known that
appearance of infrared renormalon singularities in the perturbative
calculations indicate an importance of non-perturbative higher twist effects
[14]. A nature of the non-perturbative higher twist corrections is process
dependent [9,10,14]. In the case of the pion form factor they, probably, may
be associated with contributions from the pion's higher Fock state wave
functions $\mid q\overline{q}^{\prime }g>$, etc.

\bigskip

{\bf 4.}   Some comments are in order concerning these results.
First of all, it is instructive to compare sources of the infrared
renormalons in our example and in other QCD processes considered in [3,4].
In these articles the running coupling constant approach was used in
one-loop order calculations for resummation any number of fermion bubble
insertions in the gluon propagator. This technique corresponds to partial
resummation of the perturbative series for a quantity under consideration.
In fact, in pQCD an infrared-safe and renormalization scheme invariant
quantity has the perturbative series in $\alpha _{S}(\mu ^{2})$ similar to
that shown in (5). But now the coefficients $S_{n}$ of such series have the
following expansion in powers of $\beta _{0}$%
\begin{equation}
S_{n}=c_{n}\beta _{0}^{n-1}+r_{n}\beta _{0}^{n-2}+...
\end{equation}
The Borel transform of such perturbative series defined as in (9) (that is,
using only $c_{n}$ from (19)) corresponds to partial resummation of the
series (5) and the only source of terms of order $\alpha _{S}^{n}\beta
_{0}^{n-1}$ in (5) is the running coupling constant [3,4].)

We have emphasized in Sect.2 that in our calculations we use only the
leading order term $\alpha _{S}(\widehat{Q}^{2})T_{H}^{1}$. Therefore, it is
not accidental that $S_{n}$ in (7),(8) have exactly $\sim \beta _{0}^{n-1}$
dependence and, hence, the expressions (16),(17) are exact sums of the
corresponding series. The source of term $\sim \alpha _{S}^{n}\beta
_{0}^{n-1}$ in (5) is the running coupling $\alpha _{S}(Q^{2}(1-x)(1-y))$,
but here it runs due to the integration in (1) over the pion quark's
(antiquark's) longitudinal fractional momentum. May in our case terms $\sim
\alpha _{S}^{n}\beta _{0}^{n-1}$ be also described by fermion vacuum
polarization insertion into the exchanged between the pion constituents
gluon propagator or not, require further investigations.

\bigskip

{\bf 5.}   It is useful to find how much the resummed expressions
(16),(17) differ from the ''bare'' ones. In $Figs.1,2,$ the ratio $%
R_{w.f.}=\left[ Q^{2}F_{\pi }(Q^{2})\right] _{w.f.}^{res}/$ $\left[
Q^{2}F_{\pi }(Q^{2})\right] _{w.f.}^{0}$, where $\left[ Q^{2}F_{\pi
}(Q^{2})\right] _{w.f.}^{0}$ is the pion form factor in the frozen coupling
approximation, is depicted. As is seen, the contributions of the infrared
renormalons to $\left[ Q^{2}F_{\pi }(Q^{2})\right] _{w.f.}^{res}$ are
considerable for both wave functions. Nevertheless, in both cases these
contributions can be hidden into the scale of $\alpha _{S}(\widehat{Q}^{2})$
in $\alpha _{S}(\widehat{Q}^{2})T_{H}^{1}(x,y,Q^{2})$. Indeed, let us look
for a function $f(Q^{2})=c_{1}+c_{2}\alpha _{S}(Q^{2})$, such that the
leading order term $\alpha _{S}(\widehat{Q}^{2})T_{H}^{1}(x,y,Q^{2})$ with $%
\widehat{Q}^{2}=e^{f(Q^{2})}Q^{2}$ will give for $Q^{2}F_{\pi }(Q^{2})$
approximately the same results as Eqs.(16),(17). Numerical fitting allows us
to get ($n_{f}=3$), 
\begin{eqnarray}
f(Q^{2}) &\simeq &-5.6+10.91\alpha _{S}(Q^{2}),~~~for~~ \phi _{asy},
\nonumber \\
f(Q^{2}) &\simeq &-11.76+38.09\alpha _{S}(Q^{2}),~~~for~~ \phi_{CZ}.
\end{eqnarray}
It is worth noting that a dependence of $f(Q^{2})$ on $\alpha _{S}(Q^{2})$
was introduced and argued in the next-to-leading order scale-setting
calculations in [2].

The results of numerical calculations of $R_{w.f.}$ with redefined scale in
the leading order expression are shown in $Figs.1,2.$ As is seen from $%
Figs.1,2,$ $R_{w.f.}\approx 1$ in the whole range of $Q^{2}$.

The redefinition of the scale $Q^{2}\rightarrow e^{f(Q^{2})}Q^{2}$ means,
that an average virtuality of the exchanged gluon in the pion form factor
diagrams equals to $e^{f(Q^{2})}Q^{2}$. The coupling constant $%
\alpha _{S}(e^{f(Q^{2})}Q^{2})$ for the functions under consideration (6)
varies from $\alpha _{S}\approx 0.39$ at $Q^{2}=2~GeV^{2}$ to $\alpha
_{S}\approx 0.49$ at $Q^{2}=20~GeV^{2}$ for $\phi _{CZ}$, and from $\alpha
_{S}\approx 0.54$ at $Q^{2}=2~GeV^{2}$ to $\alpha _{S}\approx 0.35$ at $%
Q^{2}=20~GeV^{2}$ for $\phi _{asy}$. These values are typical for such kinds
of studies [3,4].

It is important to notice that the functions (20) cannot be obtained
directly from perturbative series (5), (7), (8) by means of the BLM
next-to-leading order scale-setting prescription [2].

\bigskip

{\bf 6.}   In this letter we have found the resummed expression for
the pion electromagnetic form factor. For regularization of the infrared
renormalons induced divergences in the perturbative series for $F_{\pi
}(Q^{2})$ we have used the method of the Borel transform and the principal
value prescription. We have demonstrated that the resummed result depends on
the pion model wave functions used in calculations. Our investigations of
the infrared renormalon effects in the pion electromagnetic form factor
calculations allow us to conclude that exclusive processes have two sources
of infrared renormalon contributions;\\
1) integration in an exclusive process amplitude over a hadron constituents' 
longitudinal momenta,\\
2) the running coupling constant in higher-order Feynman diagrams.

The self-consistent treatment of these two sources of the infrared
renormalon effects, as well as their study by taking into account a
dependence of $\phi _{\pi }(x,\widehat{Q}^{2})$ on $\widehat{Q}^{2}$, are
subject of forthcoming publications.
\newpage

{\Large \bf REFERENCES}\vspace {10mm}\\
{\bf  1.}   S.J.Brodsky, P.G.Lepage and P.B.Mackenzie, Phys.Rev.D28 (1983)
228.\\
{\bf 2.}   S.J.Brodsky and H.Lu, Phys.Rev.D51 (1995) 3652. \\
{\bf 3.}   M.Neubert, Phys.Rev.D51 (1995) 5924.\\
{\bf 4.}   P.Ball, M.Beneke and V.M.Braun, Phys.Rev.D52 (1995) 3929;\\
P.Ball, M.Beneke and V.M.Braun, Nucl.Phys.B452 (1995) 563;\\
C.N.Lovett-Turner and C.J.Maxwell, Nucl.Phys.B452 (1995) 118. \\
{\bf 5.}  S.S.Agaev, Phys.Lett.B360 (1995) 117; E: Phys.Lett.B369 (1996)
379.\\
{\bf 6.}  S.S.Agaev, Mod.Phys.Lett.A10 (1995) 2009. \\
{\bf 7.}  G.P.Lepage and S.J.Brodsky, Phys.Lett.B90 (1979) 355;
Phys.Rev.D22 (1980) 2157;\\
A.Duncan and A.H.Mueller, Phys.Lett.B90 (1980) 159; Phys.Rev.D21
(1980) 1636;\\
A.V.Efremov and A.V.Radyushkin, Phys.Lett.B94 (1980) 245.\\
{\bf 8.}  R.D.Field, R.Gupta, S.Otto and L.Chang, Nucl.Phys.B186 (1981)
429. \\
{\bf 9.}  A.H.Mueller, Nucl.Phys.B250 (1985) 327. \\
{\bf 10.} H.Contopanagos and G.Sterman, Nucl.Phys.B419 (1994) 77. \\
{\bf 11.} V.L.Chernyak and A.R.Zhitnitsky, Phys.Rep.112 (1984) 173. 
\\
{\bf 12.} V.I.Zakharov, Nucl.Phys.B385 (1992) 452. \\
{\bf 13.} A.Erdelyi, Higher transcendental functions, v.2,
McGraw-Hill Book Company, New York, 1953. \\
{\bf 14.} I.I.Bigi, M.A.Shifman, N.G.Uraltsev and A.I.Vainstein,
Phys.Rev.D50 (1994) 2234;\\
A.V.Manohar and M.B.Wise, Phys.Lett.B344 (1994) 407;\\
B.R.Webber, Phys.Lett.B339 (1994) 148;\\
G.P.Korchemsky and G.Sterman, Nucl.Phys.B437 (1995) 415;\\
R.Akhoury and V.I.Zakharov, Saclay preprint SACLAY-SPhT-95/043,\\
April 1995.\newpage

{\Large \bf FUGURE\ CAPTIONS}\vspace{10mm}\\
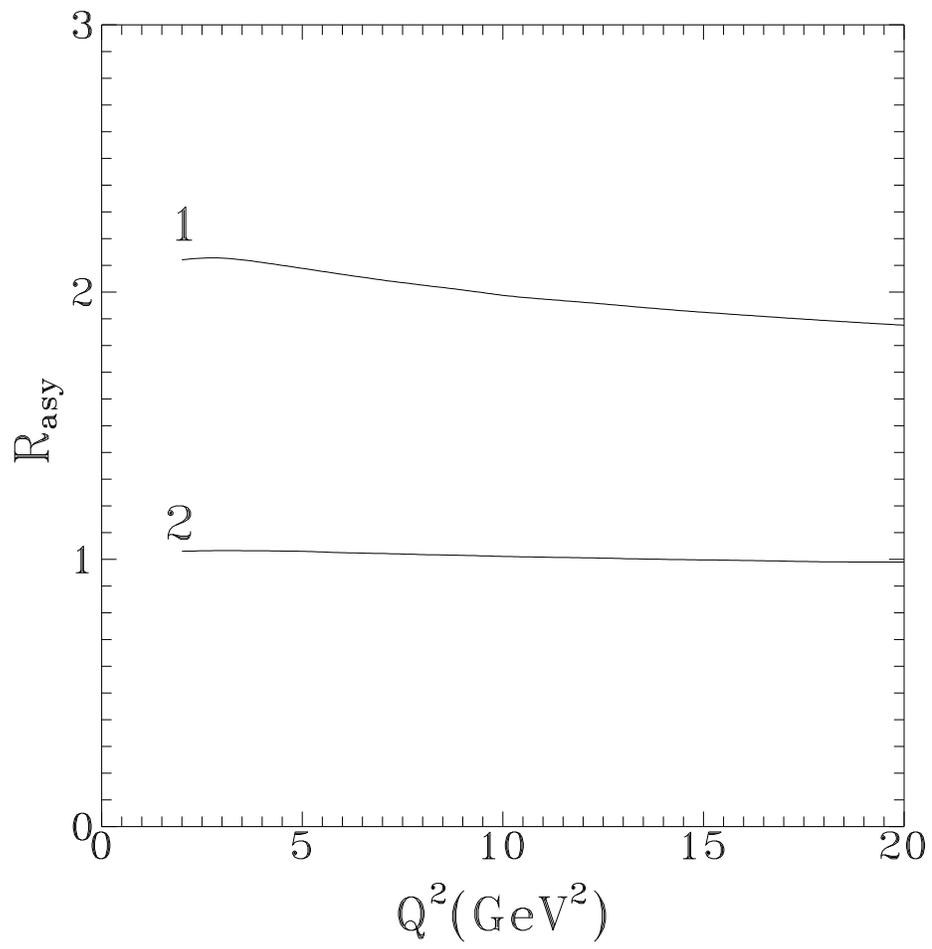
Fig.1. The dependence of the ratio $R_{asy}$ on $Q^{2}$. Here the curves 1
and 2 correspond to $R_{asy}$ with $[Q^{2}F_{\pi }(Q^{2})]_{asy}^{0}$ in the
frozen coupling constant approximation and after scale-setting procedure $%
Q^{2}\rightarrow e^{f(Q^{2})}Q^{2}$, respectively. In calculations the QCD
scale parameter $\Lambda $ has been taken equal to $0.1~GeV.$\vspace{3mm}\\
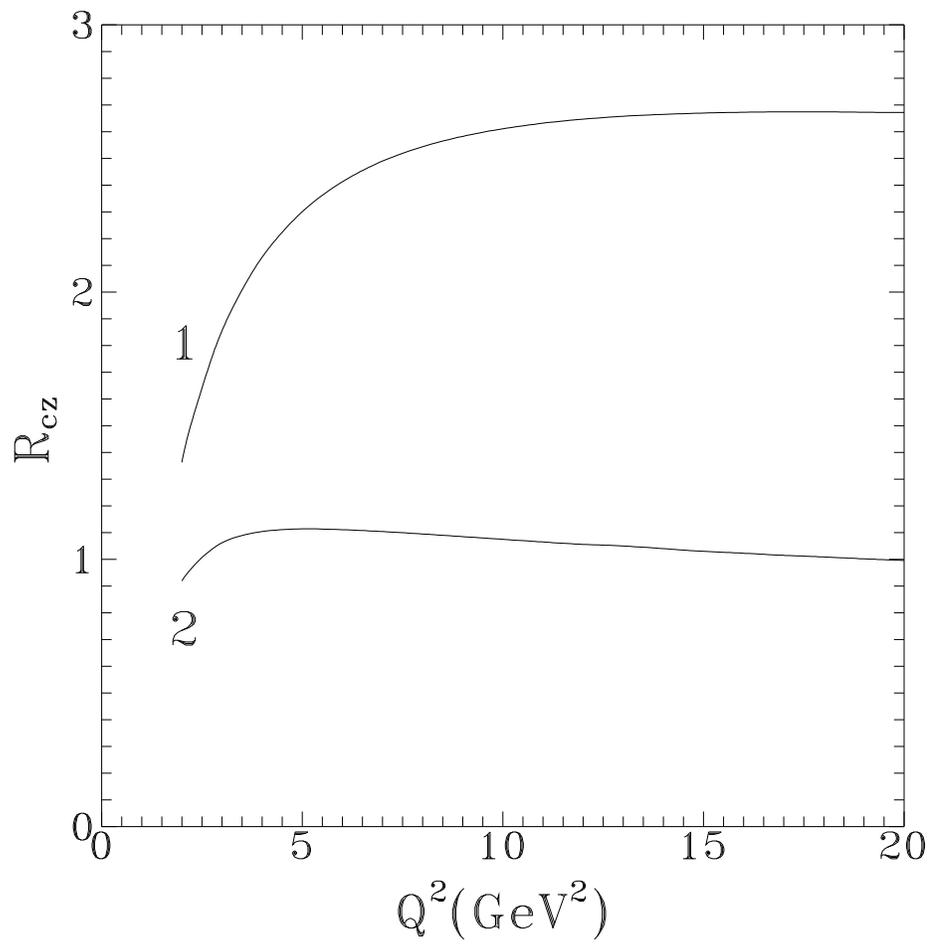
Fig.2. The same as in Fig.1, but for $R_{CZ}.$

\end{document}